\documentclass[sn-mathphys-num]{sn-jnl}% Math and Physical Sciences Numbered Reference Style 
\usepackage{graphicx}%
\usepackage{multirow}%
\usepackage{amsmath,amssymb,amsfonts}%
\DeclareFontFamily{U}{stix2bb}{}
\DeclareFontShape{U}{stix2bb}{m}{n} {<-> stix2-mathbb}{}

\NewDocumentCommand{\indicator}{}{\text{\usefont{U}{stix2bb}{m}{n}1}}
\DeclareMathOperator*{\argmax}{arg\,max}

\usepackage{amsthm}%
\usepackage{mathrsfs}%
\usepackage[title]{appendix}%
\usepackage{xcolor}%
\usepackage{textcomp}%
\usepackage{manyfoot}%
\usepackage{booktabs}%
\usepackage[ruled]{algorithm2e}
\usepackage{listings}%
\usepackage{subfigure}
\usepackage{graphicx,subcaption}
\theoremstyle{thmstyleone}%
\theoremstyle{thmstyletwo}%

\theoremstyle{thmstylethree}%

\begin{document}

\title[Article Title]{Realizing Quantum Adversarial Defense on a Trapped-ion Quantum Processor}


\author*[1]{\fnm{Alex} \sur{Jin}}\email{e1051010@u.nus.edu}

\author[1,2]{\fnm{Tarun} \sur{Dutta}}\email{tarunduttaz@gmail.com}
\equalcont{These authors contributed equally to this work.}

\author[3]{\fnm{Anh Tu} \sur{Ngo}}\email{ngoanhtu001@e.ntu.edu.sg}
\equalcont{These authors contributed equally to this work.}

\author[3]{\fnm{Anupam} \sur{Chattopadhyay}}\email{anupam@ntu.edu.sg}
\equalcont{These authors contributed equally to this work.}

\author[1,4]{\fnm{Manas} \sur{Mukherjee}}\email{cqtmukhe@nus.edu.sg}
\equalcont{These authors contributed equally to this work.}

\affil*[1]{\orgdiv{Centre for Quantum Technologies}, \orgname{National University of Singapore}, \orgaddress{\street{3 Science Dr 2}, \city{Singapore}, \postcode{117543}, \country{Republic of Singapore}}}

\affil[2]{\orgdiv{School of Physics}, \orgname{University of Hyderabad}, \country{India}, \postcode{500046}}

\affil[3]{\orgdiv{College of Computing and Data Science (CCDS)}, \orgname{Nanyang Technological University}, \orgaddress{\street{50 Nanyang Ave, Block N 4}, \city{Singapore}, \postcode{639798}, \country{Republic of Singapore}}}

\affil[4]{\orgdiv{Institute of Material Research and Engineering (IMRE)}, \orgname{Agency for Science, Technology and Research (A*STAR)}, \orgaddress{\street{2 Fusionopolis Way, Innovis \#08-03}, \city{Singapore}, \postcode{138634}, \country{Republic of Singapore}}}


\abstract{Classification is a fundamental task in machine learning, typically performed using classical models. Quantum machine learning (QML), however, offers distinct advantages, such as enhanced representational power through high-dimensional Hilbert spaces and energy-efficient reversible gate operations. Despite these theoretical benefits, the robustness of QML classifiers against adversarial attacks and inherent quantum noise remains largely under-explored. In this work, we implement a data re-uploading-based quantum classifier on an ion-trap quantum processor using a single qubit to assess its resilience under realistic conditions. We introduce a novel convolutional quantum classifier architecture leveraging data re-uploading and demonstrate its superior robustness on the MNIST dataset. Additionally, we quantify the effects of polarization noise in a realistic setting, where both bit and phase noises are present, further validating the classifier’s robustness. Our findings provide insights into the practical security and reliability of quantum classifiers, bridging the gap between theoretical potential and real-world deployment.}
% \abstract{Classification of data is a fundamental process in machine learning, typically performed using classical models. Quantum machine learning (QML) models, however, offer distinct advantages, including higher-dimensional representational power and energy-efficient reversible gate operations. Despite these theoretical benefits, the practical robustness of QML classifiers against adversarial attacks and inherent noise remains underexplored. Here, we implement a data re-uploading-based quantum classifier on an ion-trap quantum computer with a single qubit to evaluate its resilience against these challenges. We propose a novel convolutional quantum classifier architecture leveraging data re-uploading and demonstrate its superior robustness on the MNIST dataset. Additionally, we quantify the impact of polarization noise in a realistic setting where both bit and phase noise are present, further validating the classifier’s robustness. Our findings provide insights into the practical security and reliability of quantum classifiers, advancing their potential applicability in real-world scenarios.}




\maketitle

\section{Introduction}\label{sec1}

Classification is a fundamental aspect of learning, present in both natural cognition and artificial intelligence. For example, children learn to classify food preferences, and traders distinguish shares to buy or sell. The data obtained from real-life situations, such as images and soundtracks, is often noisy. Yet, modern classical machine learning (ML) models can classify data into multiple classes with nearly 100\% accuracy. The complexity and dimensionality of data often require the exploration of higher-dimensional hyperspaces in order to achieve better class separation. Quantum machine learning offers an alternative approach by naturally leveraging the large-dimensional Hilbert spaces of quantum systems. Another potential advantage of quantum mechanics is the ability to explore Hilbert spaces more efficiently through superposition and unitary/reversible evolution, which could allow QML classifiers to achieve comparable or superior classification performance with fewer computational resources and reduced energy consumption. Despite the success of classical ML-based classifiers, they remain prone to certain noise and distortions that may occur in realistic data or could be injected to compromise the accuracy of a classifier through adversaries. The latter, also known as an adversarial attack, is a concern of this article. \\

This is a nascent but fast-growing research field, mostly taking cues from classical classifiers. In the case of classical classifiers, certain strategies perform better than others depending on the classification task. However, there is no universal strategy that provides both high robustness and efficacy on adversarial datasets. Quantum machine learning (QML) offers new resources, such as superposition, compared to its classical counterpart. Therefore, some of the valid research questions to ask are: Can QML provide inherent robustness against adversarial attacks? Will the cost of training a quantum model against adversarial attacks be lower? Finally, can we quantify the robustness of real-world quantum classifiers?  \\

In order to study this problem, we must first choose a model and a corresponding algorithm for the classification task. Below, we justify the choice of model for this study. In quantum classifiers, regardless of the algorithm, parameter optimization is performed on a classical computer. Consequently, these algorithms are variational quantum-classical hybrid algorithms \cite{mcclean2016theory}. On noisy intermediate-scale quantum (NISQ) computing hardware, variational-type hybrid algorithms are believed to offer practical advantages over fully quantum algorithms like Shor’s factorization algorithm \cite{shor1999polynomial}. Quantum classifier algorithms are broadly categorized into three types: explicit, implicit, and data re-uploading~\cite{perez2020data}, based on how classical data and optimization parameters are represented \cite{jerbi2023quantum}. Significant progress has been made towards establishing a unified framework for all quantum classifiers \cite{Schuld2021}. Among these, explicit algorithms are the most studied; according to the \textit{Representer Theorem} \cite{scholkopf2001generalized}, they guarantee superior training accuracy with the same training set compared to implicit ones.\\

Two recent theoretical advancements are particularly relevant to our discussion. First, a unified framework has been developed to compare the performance and resource requirements of all three quantum classifier types. Second, a robustness quantification method, inspired by differential privacy in classical computing, has been proposed---leveraging quantum depolarizing noise for masking. While these theoretical insights provide valuable benchmarks, experimental validation of these findings on NISQ hardware remains scarce.

The data re-uploading algorithm (DRA) shows promise for achieving provable quantum advantages in training efficiency and reduced data requirements \cite{jerbi2023quantum}. Previous experimental work by our group demonstrated that DRA matches the classification accuracy of classical neural networks with comparable resources. Our previous experimental studies confirmed that DRA can achieve classification accuracy comparable to classical neural networks while using similar computational resources. Notably, we demonstrated that DRA enables autonomous training without reliance on classical simulations, a significant advantage over most existing variational algorithm implementations\cite{dutta2024trainability}.

Here, we extend previous research by implementing adversarial attacks on quantum classifiers, demonstrating that such attacks can significantly degrade classification accuracy---much like in classical machine learning. To address this vulnerability, we also propose and experimentally demonstrate a new quantum classifier approach that exhibits higher robustness against such attacks. We quantify the robustness of our solution using depolarization noise, following the methodology outlined in ref. \cite{weber2021optimal}. As mentioned in much of the literature in this field, applications on real quantum hardware remain rare and often a predicament to improve on the models \cite{ren2022experimental}. \\

In a real NISQ device, noise sources extend beyond depolarization to include bit and phase flips. By accounting for these effects, our work provides a more comprehensive evaluation of the robustness and potential quantum advantage of our proposed quantum classifier against adversarial attacks in a practical setting. As in classical machine learning, no universal solution exists for defending against all adversarial attacks. However, our approach serves as a foundation for further training on such datasets to improve robustness. In terms of quantifying robustness, the currently proposed hypothesis, based on classical classifiers, largely aligns with the QML model. However, in realistic scenarios, a detailed understanding of the error budget is essential. \\     

\begin{figure}
\centering     %%% not \center
% \subfigure[QAML]{
\subfigure{
\includegraphics[width=0.9\textwidth]{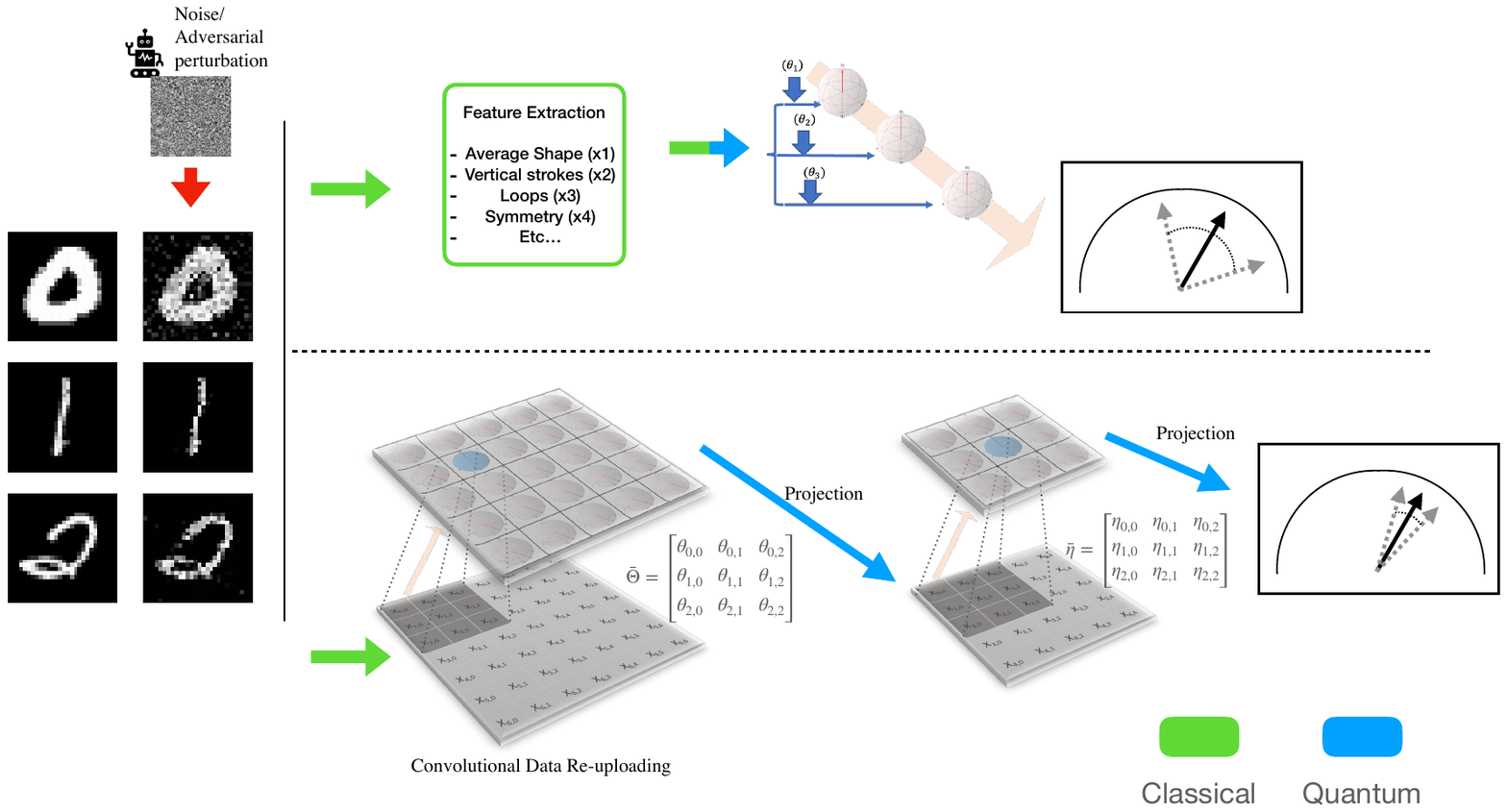}}
% \caption{QAML architecture. \textit{Left}: examples of image data and their variations under noise or adversarial perturbations. \textit{Top right}: DRA architecture based on classical feature extractors. Input images are first processed classically into a low dimensional feature space, and these features are re-uploaded along with trainable variational parameters to a quantum classifier. \textit{Bottom right}: DRA-CQC architecture without any classical feature extraction. Raw pixel values are convoluted with trainable parameters and processed via the DRA. The DRA-CQC architecture demonstrates higher resilience against noise and stronger robustness against adversarial perturbation.}
\caption{QAML architecture. \textit{Left}: Examples of image data with variations due to noise or adversarial perturbations. \textit{Top right (DRA Architecture)}: A hybrid quantum-classical model where input images are first processed by a classical feature extractor, reducing them to a low-dimensional feature space. These extracted features are re-uploaded along with trainable variational parameters to a quantum classifier. \textit{Bottom right (DRA-CQC Architecture)}: A fully quantum model that directly processes raw pixel values without classical feature extraction. Trainable parameters are applied through convolution, and the data is processed via DRA. This method shows greater resilience to noise and stronger robustness against adversarial attacks.}

\label{fig:qaml}
\end{figure}

In the following, we first present our results on the successful design of an efficient method for generating adversarial datasets based on the original MNIST data. We then describe our initial attempt to counter this attack using a simplistic QML strategy. We then introduce our novel  convolutional quantum classifier (CQC) architecture, and demonstrate its effectiveness in mitigating these adversarial attacks. A detailed comparison between the two approaches is provided in the methods section. Finally, we benchmark our strategy using a quantifiable definition of robustness, considering only depolarization noise. To our knowledge, this is the first experimental measurement of quantum classifier robustness under adversarial attacks. Our analysis establishes a method for distinguishing different noise components in quantum systems, extending beyond depolarization noise. In Fig. \ref{fig:qaml}, we outline the overall architecture of our inference architectures as well as specifying the injection of adversarial and noisy data.

\section{Results}\label{sec2}

The data re-uploading algorithm is a promising algorithm in the general explicit quantum classifier family. Mathematically, a data re-uploading model defines a mapping $f: \mathscr{X} \rightarrow \mathscr{F} \rightarrow \mathscr{Y}$, where $\bar{x}\in \mathscr{X}$ are vectors of classical data in $\mathbb{R}^n$, $\mathscr{F}$ is the quantum feature space mapped using $\rho_{\bar{\theta}}(\bar{x})$, and $\bar{y} \in \mathscr{Y}$ are m-dimensional vectors of the output space. The resulting composite map can be expressed as expectation values of the following form:

\begin{align}
    f_{\bar{\theta}}(\bar{x}) = \mathrm{Tr}[\rho_{\bar{\theta}}(\bar{x})O_{\bar{\theta}}],
    \label{equation:dra-expectation}
\end{align}

where $\bar{\theta}$ are the variational parameters of our circuit, and $\rho_{\bar{\theta}}(\bar{x}) = U(\bar{x},\bar{\theta})|0\rangle\langle0|U^\dag(\bar{x},\bar{\theta})$. In general, $U$ consists of $L$ parameterized layers in the form of $U(\bar{x},\bar{\theta})=\prod_{l=1}^L U_l(\bar{\theta}_l, \bar{x})$, and its precise definition is called the \textit{ansatz} of a model. Finally, we have an (variational) observable $O_{\bar{\theta}}=V_L(\bar{\theta})^\dag O V_L(\bar{\theta})$.

In a recent study, it has been shown that all parameterized quantum circuits fall under the general umbrella of linear quantum models, and that data re-uploading models are \textit{exponentially} more resource efficient in terms of the number of qubits and training data points\cite{jerbi2023quantum}. Therefore, data re-uploading algorithm is a natural candidate for exploring QML algorithms, particularly in the NISQ-era. We will first introduce the results obtained from two approaches used for the classification of the MNIST data: (a) principal component analysis followed by DRA and (b) convolutional quantum classifier using DRA.   

% \subsection{Kernel methods}

\begin{table}[h]
\caption{Train/Test Accuracy, Noise Resilience, and Perturbation Robustness}
\begin{tabular}{@{}lllll@{}}
\toprule
 & & Binary (0,1)  & Multiclass (0,1,2) & Multiclass (0,1,2,3)\\
\midrule
\textbf{PCA-based} & Preprocessor & linear PCA & kernel PCA & - \\
\cmidrule{3-5}
\textbf{Quantum} & Trainable parameters & 9 & 9 & - \\
\cmidrule{3-5}
\textbf{Classifier} & Training accuracy & 100\% & 95.53\% & - \\
 & (simulation) & & & \\
\cmidrule{3-5}
 & Test accuracy & 99.61\% & 93.32\% & - \\
 & (simulation) & & & \\
\cmidrule{3-5}
 & Test accuracy & \textcolor{blue}{99.52\%}\footnotemark[2] & \textcolor{blue}{64.05\%}\footnotemark[2] & - \\
 & (quantum computer) & & & \\
\cmidrule{3-5}
 & Adversarial accuracy & 45.14\% & 48.00\% & - \\
 & (simulation) & & & \\
 \cmidrule{3-5}
 & Adversarial accuracy & \textcolor{blue}{45.24\%}\footnotemark[2] & \textcolor{blue}{44.00\%}\footnotemark[2] & -\\
 & (quantum computer) & & & \\
\midrule
\textbf{DRA-CQC}\footnotemark[1] & Preprocessor & - & - & - \\
\cmidrule{3-5}
 & Trainable parameters & 90 & 108 & 135 \\
\cmidrule{3-5}
 & Training accuracy & 99.16\% & 94.97\% & 92.49\% \\
 & (simulation) & & & \\
\cmidrule{3-5}
 & Test accuracy & 98.28\% & 94.10\% & 91.64\% \\
 & (simulation) & & & \\
 \cmidrule{3-5}
 & Test accuracy & \textcolor{cyan}{92.00\%}\footnotemark[3] & \textcolor{blue}{93.49\%}\footnotemark[2] & \textcolor{blue}{85.81\%}\footnotemark[2] \\
 & (quantum computer) & & & \\
\cmidrule{3-5}
 & Adversarial accuracy & 90.00\% & 82.00\% & -\\
 & (simulation) & & & \\
 \cmidrule{3-5}
 & Adversarial accuracy & \textcolor{cyan}{88.00\%}\footnotemark[3] & \textcolor{blue}{78.10\%}\footnotemark[2] & -\\
 & (quantum computer) & & & \\
\botrule
\end{tabular}
\footnotetext[1]{Data Re-uploading Algorithm-based Convolutional Quantum Classifier.}
\footnotetext[2]{\textcolor{blue}{From noisy simulation matching ion-trap quantum device.}}
\footnotetext[3]{\textcolor{cyan}{From ion-trap quantum device.}}
% \footnotetext{\textit{Table Summary} Comparison of training and inference accuracies of various quantum circuits under noiseless (simulation) and noisy (quantum computer) environments. Adversarial robustness against attacks designed specifically at each test set is also given on both noiseless and noisy environments. Individual results of the Binary classification test accuracy on our quantum hardware is shown in Fig. \ref{fig:resilience-vs-depolar} and Fig. \ref{fig:robustness-vs-adv}.}
\footnotetext{\textit{Table Summary} Comparison of Train/Test Accuracy, Noise Resilience, and Adversarial Robustness for Quantum Classifiers. This table presents the training and test accuracies of different quantum classifiers under simulation (noiseless) and real quantum hardware (noisy) conditions for binary and multiclass classification tasks. Adversarial robustness is evaluated against attacks specifically designed for each test set in both noiseless (simulation) and noisy (quantum hardware) environments. Individual results of the Binary classification test accuracy on our quantum hardware are shown in Fig. \ref{fig:resilience-vs-depolar} and Fig. \ref{fig:robustness-vs-adv}.}
\label{tab:training-and-inference}
\end{table}

\subsubsection{Performance of PCA and CQC}

An alternative approach for embedding high-dimensional classical data into a quantum circuit involves dimensionality reduction techniques \cite{van2009dimensionality}. Among these, principal component analysis (PCA) is a widely utilized tool which we call as the simplistic quantum classifier. In this approach, classical datasets, such as MNIST, are first projected onto a lower-dimensional feature space. These compressed feature vectors are then mapped onto a data re-uploading architecture (DRA) by linearly combining the features with a vector of trainable parameters, with the resulting values used to parameterize quantum gates. To enhance classification accuracy, the data is re-uploaded multiple times using different sets of trainable parameters before performing quantum measurements to classify the input image. Intuitively, the efficacy of this architecture is linked to the quality of separation achieved by the classical feature extraction process. We refer interested readers to \cite{perez2021one, perez2020data} for the theoretical foundations and experimental implementation of such DRA architectures. On the other hand, the Data Re-uploading Convolutional Quantum Classifier (DRA-CQC) architecture, the details of it is explained in \ref{sec:CQC-arch}, relies on the successive convolution of the pixels of an image. In the DRA-CQC architecture, our NISQ device takes as input raw pixel values of the image patches and outputs an array of probabilities of the predicted classes.\\

We can see from table~(\ref{tab:training-and-inference}) that we are able to achieve reasonable training and test accuracy for both the binary and 3-class problems from the MNIST dataset using PCA-based DRA. However, such a classification schema is susceptible to perturbation from two perspectives. First, as shown in the table, we notice a significant drop in test accuracy from noiseless simulation to noisy simulation of our NISQ device for the 3-class problem. Second, there is also a drop in both the training and test accuracies even for noiseless simulation when the problem is related to higher number of classes, binary to 3-class. On the contrary, the DRA-CQC approach demonstrates strong robustness against both noise and increased classification complexity. Even when extending to the 4-class problem, noise in the system reduces test accuracy by only 6\%. We believe the robustness lies in the averaging of the noise due to the convolution. A more quantitative assessment of robustness is provided by dedicated robustness measurements. While test accuracy serves as a general indicator of an algorithm’s resistance to arbitrary noise, it does not effectively measure robustness against curated adversarial noise specifically designed to induce misclassification. \\

We employed a Genetic Algorithm (GA)-based adversarial image generator to produce images that closely resemble the MNIST dataset. For example, with an average pixel value perturbation of 12.6\% using an attack strength of $1.0$ for both $w_0$ and $w_1$ (see \ref{sec:methds}), these generated images were misclassified when evaluated on a trained classifier, despite their resemblance to the benign images. The classification results for both the PCA-based and DRA-CQC classifiers are presented in Table~(\ref{tab:training-and-inference}). Notably, the PCA-based classifier performs no better than random guessing, highlighting its vulnerability to adversarial perturbations. In contrast, the DRA-CQC classifier demonstrates significant robustness, exhibiting only a moderate decrease in accuracy. Specifically, for the binary classification task, accuracy drops by merely 4\%, while for the three-class classification task, the decline is limited to 14\%.

\subsection{Quantifying robustness against adversaries}

\begin{figure}
\centering     %%% not \center
\subfigure{
\includegraphics[width=0.99\textwidth]{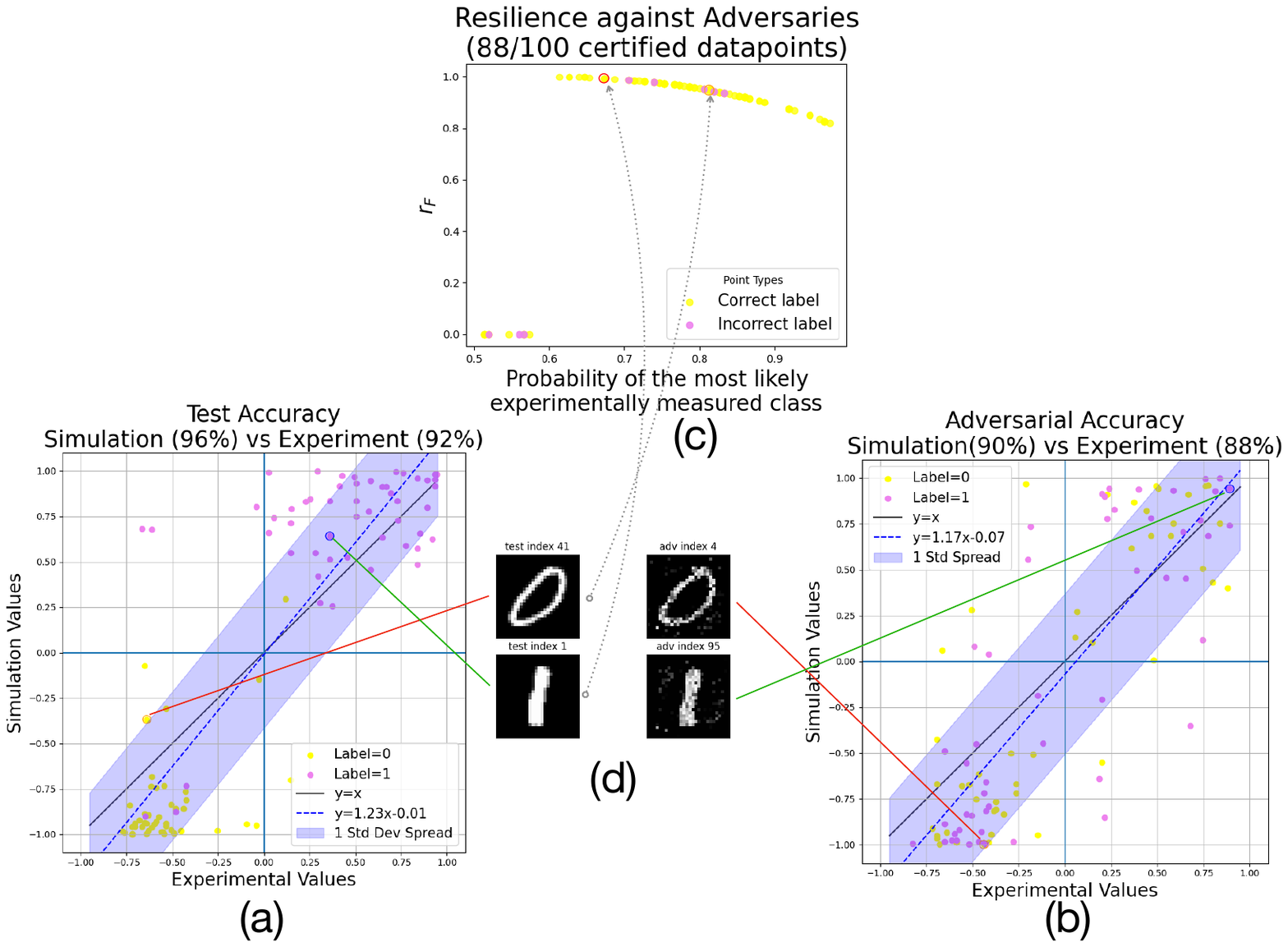}}
\caption{Resilience against adversaries. \textbf{a}. Predicted values of a trained DRA-CQC on an MNIST test set consisting of 100 images of handwritten figures of 0s and 1s. The decision boundaries for predicted value from our NISQ device and numerical simulation are the y- and x- axes, respectively. The blue band covers the 1 standard deviation of experimental uncertainty benchmarked against the simulation outputs. \textbf{b}. Similar as a, but the MNIST test set is adversarially attacked. \textbf{c}. Resilience against adversaries of all 100 test set images certified at probability 0.9. $r_F$ at 0 represent uncertified data points. The colors are added to illustrate whether our experimental measurements produced correct labels. \textbf{d}. Two example benign input images used in a and two example adversarial input images (closest to the benign images under the $L_2$-norm) used in b, with their corresponding predicted values circled in the previous plots.}
\label{fig:robustness-vs-adv}
\end{figure}

In security-sensitive domains, such as autonomous vehicle decision-making and medical data classification, quantifying the robustness of machine learning systems against adversarial interference is critically important. This entails analyzing the model's behavior under intentional input perturbations, typically constrained by a fidelity threshold of $1-\epsilon$. In classical machine learning, this robustness is often quantified using the concept of certified accuracy at a given radius r, where $L_p$-norms are commonly used to measure the magnitude of perturbations. Adversarial robustness, a key metric in this context, evaluates the ability of a trained model to resist manipulation by adversarial attacks, ensuring accurate predictions even under deliberate data alterations. More precisely, given a trained classical machine learning model $f:\mathbb{R}^n \rightarrow \mathscr{K}$, mapping n-dimensional inputs into K distinct output classes, the model is said to be certified at radius $\epsilon$ if its output classes on the test set remain unchanged when the input is perturbed by at most $\epsilon$, usually measured in terms of $l_0$-, $l_2$-, or $l_{\infty}$-norms.

For QML classifiers, an analogous definition has been given in \cite{weber2021optimal} as the following. Given a set of labeled test data $\mathscr{T} = \{(\sigma_i, y_i)\}_{i=1}^{|\mathscr{T}|}$, the certified test set accuracy at fidelity $1-\epsilon$ is defined as

\begin{align}
    \frac{1}{|\mathscr{T}|} \sum_{(\sigma,y)\in\mathscr{T}} \indicator \{\mathbb{A}(\sigma)=y\land r_F(\sigma)\leq 1-\epsilon\},
    \label{equation:dra-expectation}
\end{align}

where $\sigma$ represents a quantum state, $y$ represents a classical label, $\indicator$ is an indicator function, and $\mathbb{A}$ is a quantum classifier. $r_F$ is the minimum robustness fidelity defined as $r_F = \frac{1}{2}(1+\sqrt{1-p_B-p_A(1-2p_B)+2\sqrt{p_Ap_B(1-p_A)(1-p_B)}})$, where $1\geq p_A\geq p_B\geq 0$, are the highest two values of the classifier's output probability vector.\\

We now show a concrete experimental verification of this robustness quantifier using the DRA-CQC architecture. The details of this architecture is explained in \ref{sec:CQC-arch}. In the DRA-CQC architecture, our NISQ device takes as input raw pixel values of the image patches and outputs an array of probabilities of the predicted classes. In Fig. \ref{fig:robustness-vs-adv}, we illustrate both simulated and predicted output probabilities on the binary classification task of predicting a subset of 0s and 1s from the MNIST dataset. We also give the minimum robustness fidelity, $r_F$, as defined in \ref{equation:dra-expectation}, of each of the predicted outputs. As an interesting sidenote, a quick tally shows that $25\%$ (3 out of 12) of the uncertified points are incorrectly classified where as $5.7\%$ (5 out of 88) of the certified points are incorrectly labeled, as shown in Fig. \ref{fig:robustness-vs-adv}c.

\subsection{Robustness against depolarization noise}

\begin{figure}
\centering     %%% not \center
\subfigure{
\includegraphics[width=0.9\textwidth]{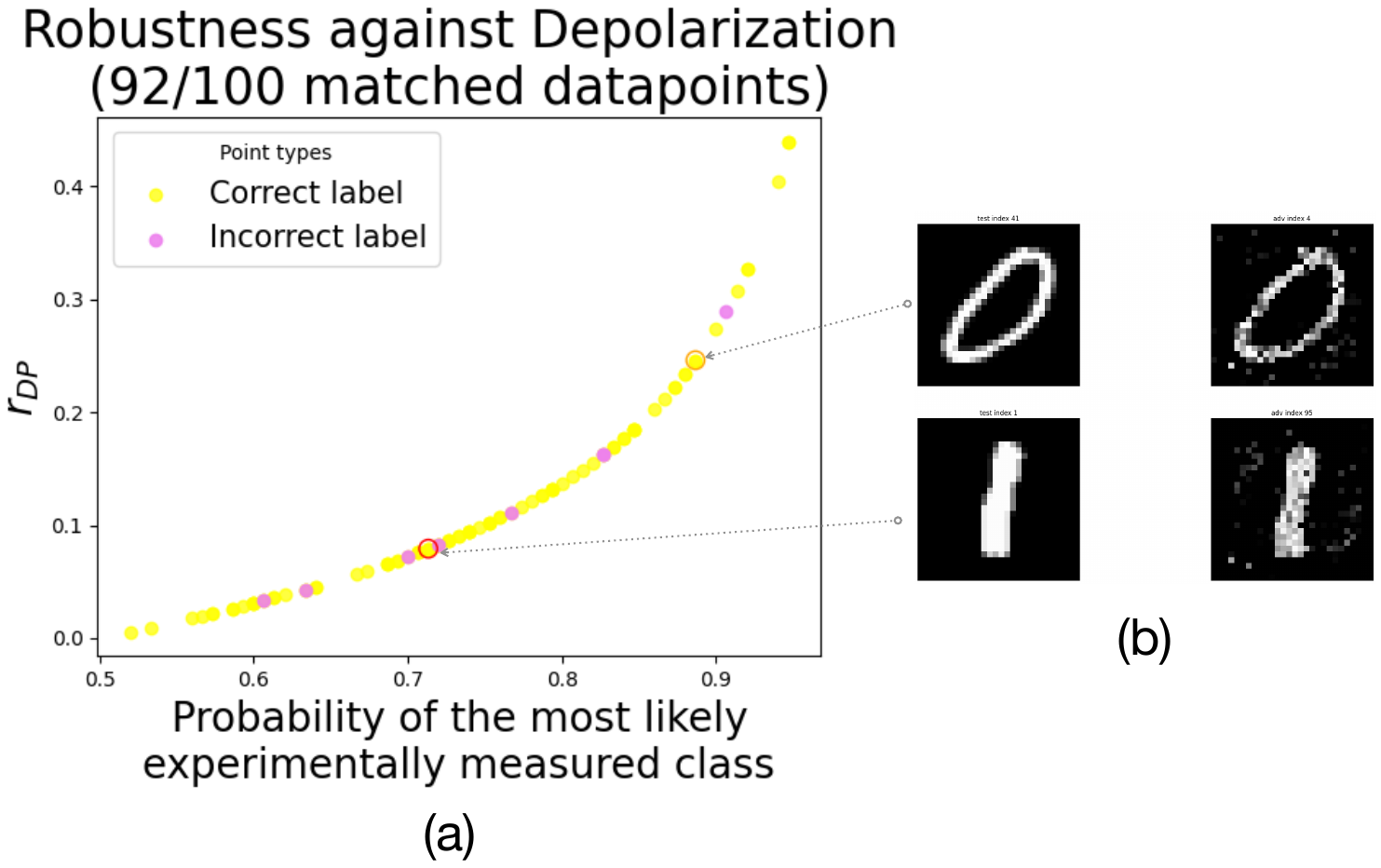}}
% \caption{Robustness against depolarization noise. \textbf{a}. Robust of all 100 test set images measured from our experimental NISQ device. The colors are added to illustrate whether our experimental measurements produced correct labels. \textbf{b}. The same example input images as used in Fig. \ref{fig:robustness-vs-adv}. Higher robustness values indicate stronger robustness.}
\caption{Robustness against depolarization noise. \textbf{a} Robustness analysis of all 100 test set images measured on our experimental NISQ device. Colors indicate whether the experimental measurements produced correct labels for classifications. \textbf{b} The same example input images as used in Fig. \ref{fig:robustness-vs-adv}. Higher robustness values correspond to stronger resistance to noise effects.}
\label{fig:resilience-vs-depolar}
\end{figure}

Depolarization noise or depolarization channel refers to a type of quantum noise induced by the environment or the system, where the coherence of a quantum state is reduced and is driven toward a completely mixed state $\frac{I}{d}$, where $I$ is the identity matrix and $d$ is the dimension of the Hilbert space. Mathematically, for a quantum state $\sigma$, depolarization can be described as:

\begin{align}
    \sigma '=(1-p)\sigma + p \frac{I}{d},
    \label{equation:depolarization-noise}
\end{align}

where $p$ refers to the depolarization probability, $p\in [0,1]$. In \cite{du2021quantum}, it has been shown that for a given quantum state $\sigma$, robustness can be guaranteed for any adversarial state $\rho$ with,

\begin{align}
    T(\rho, \sigma) < r_{DP}(p) := \frac{p}{2(1-p)}(\sqrt{\frac{p_A}{1-p_A}}-1).
    \label{equation:depolarization-noise}
\end{align}

$T(\rho, \sigma)$ is a distant metric, such as the $L_d-norm$, of the input states. We note that $p_A$ has the same definition as defined in the previous subsection. Our experimental results shown in Fig. \ref{fig:resilience-vs-depolar} show that the depolarization robustness follow a similar trend to those found in \cite{weber2021optimal}. However, the effect of other noise sources such as bit-flip and phase-flip errors also contributes to the total noise of our system and causes our findings to deviate from purely theoretical predictions.

% \subsection{Robustness of data re-uploading algorithms}

% \begin{figure}
% \centering     %%% not \center
% \subfigure[ADV]{
% \includegraphics[width=0.9\textwidth]{figures/adversarial_example.png}}
% \caption{Adversarial Example}
% \label{fig:adversarial-example}
% \end{figure}

% A major open problem of running QML algorithms on NISQ-devices is the algorithm's robustness to noise, which can be unintentional perturbations from the environment and the physical system, or due to adversarial attacks from malicious parties. Recent work \cite{weber2021optimal} has outlined a framework for quantifying the robustness of QML algorithms from a hypothesis testing point of view. Such perturbations can lead to catastrophic consequences as illustrated in \ref{fig:adversarial-example}. In the following subsections we will examine the robustness of a toy DRA classifier trained on the MNIST dataset running on our trapped-ion based quantum computer. 
% We will first study the classifier's robustness to noise from the environment and the system, and then explore its resilience against malicious attackers.

\section{Discussion}
\label{sec:discussion}

In this work, we have systematically examined the impact of both noise and adversarial attacks on a quantum classifier based on the data re-uploading model. We introduced and implemented a convolutional quantum classifier (CQC), demonstrating its enhanced robustness against adversarial perturbations on the MNIST dataset. Furthermore, we quantified the robustness conferred by depolarization noise in a real NISQ device, providing a practical measure of noise resilience in quantum classifiers. This study presents a comprehensive evaluation of the measurable robustness of the newly proposed DRA-CQC architecture, leveraging the data re-uploading algorithm in the context of MNIST classification.\\

Our findings indicate that while DRA-CQC offers an initial layer of defense against adversarial datasets, scaling quantum hardware remains a critical challenge for extending these benefits to larger, more complex datasets. Although the data re-uploading algorithm is highly resource-efficient due to the quantum universal approximation theorem \cite{jerbi2023quantum}, its hybrid quantum-classical nature makes training time-intensive in practical implementations. In terms of noise-protected robustness, our results suggest that the quantifiable advantages of depolarization noise can be leveraged in real NISQ devices, provided that other sources of noise, such as bit and phase noise, are properly mitigated. These insights underscore both the promise and the current limitations of quantum classifiers, paving the way for future advancements in scalable, adversarially robust quantum machine learning.\\

\section{Methods}
\label{sec:methds}

\subsection{Feature-based quantum classifier}

In our recent research \cite{dutta2024trainability}, we demonstrated that a quantum machine based on ion trap technology can serve as a universal quantum classifier. We employed a data re-uploading algorithm tailored to leverage the fixed Hilbert space of systems with a limited number of qubits. This alignment between the ion-trap device's capabilities and the algorithmic needs highlights the critical importance of executing highly accurate quantum operations to achieve optimal performance.

Building on the success of recent research into single-qubit quantum classifiers, we further explore the robustness of these classifiers when applied to real-world datasets.

In the data re-uploading quantum classifier paradigm, we always start the system with a qubit in the initial state $|0\rangle$. The input to our classifier are vectors $\bar{x} \in \mathbb{R}^d$, where $d$ is the dimension of the feature space. We define a sequence of unitary operations $U_l$, $1\leq l \leq L$, such that the final state $|\phi\rangle$ is:

\begin{align}
    |\phi\rangle = U_L(\bar{\theta}_L, \bar{x}) U_{L-1}(\bar{\theta}_{L-1}, \bar{x}) \dots U_1(\bar{\theta}_1, \bar{x}) |0\rangle = \prod_{l=1}^L U_l(\bar{\theta}_l, \bar{x})|0\rangle \
    \label{equation:abstract-classifier}
\end{align}

where $\vec x$ is a point from training data, and $\bar{\theta}$ is the trainable parameters we select according to the classifier's architecture. The Ansätze used here for data reuploading into the circuit defined in Eq.~\eqref{eq:ansatz_a}. 
\begin{equation}\label{eq:ansatz_a}
U(\vec\theta, \vec{x}) =  R_z(\vec{w} \cdot \vec x + b) R_y(\vec{w} \cdot \vec x + b). 
\end{equation}
 Here $R_z$ and $R_y$ are single-qubit rotations around the $z$ and $y$ axis.
Finally, we select N label states $\psi_n$, $1\leq n \leq N$, based on the number of classes we are trying to classify. The predicted class is then the label state with the largest projected population $\argmax_{n} \langle \psi_n | \phi\rangle$.

We now illustrate the application of the architecture outlined above with a binary classification problem. A natural choice of label states would be $|0\rangle$ for label 0 and $|1\rangle$ for label 1. We project our final state to the label states, $\langle 0|\phi\rangle$ and $\langle 1|\phi\rangle$, the quantum analogy of logits. The goal for the optimizer is to find a set of parameters $\bar{\theta}$ such that the final states of blue and orange data points are well separated in the Hilbert space of a single qubit.

\begin{figure}[ht]
    \centering
    \subfigure[\hspace{0mm} ]{
        \includegraphics[width=0.25\textwidth]{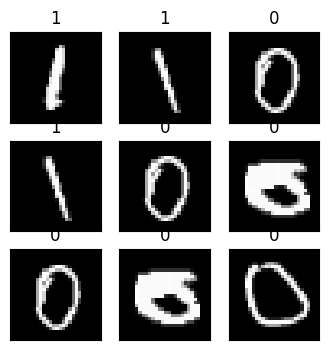}
        \label{fig:binary-dra-a}
        }
    \vspace{0 mm}
    \subfigure[\hspace{0mm} ]{
        \includegraphics[width=0.34\textwidth]{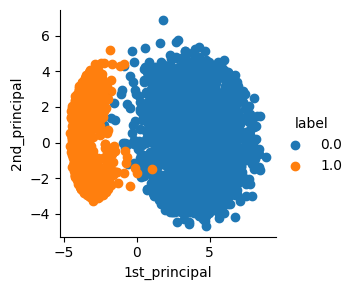}
        \label{fig:binary-dra-b}
        }
    \vspace{0 mm}
    \subfigure[\hspace{0mm} ]{
        \includegraphics[width=0.34\textwidth]{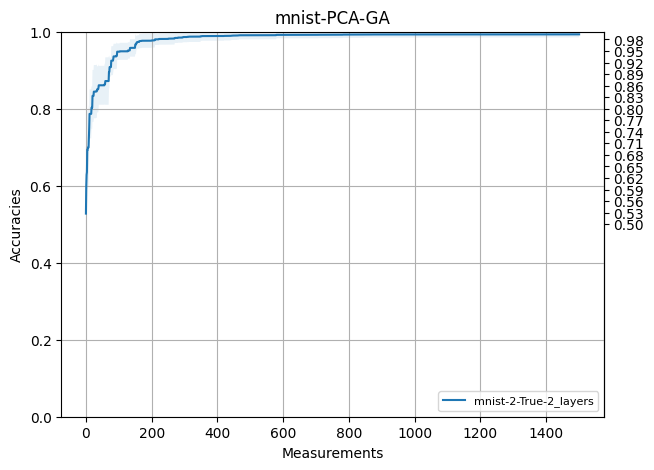}
        \label{fig:binary-dra-c}
        }
    \caption{(a) Binary classification problem: digits 0 and 1 from the MNIST dataset. (b) PCA projection: projection of (a) down to a 2 dimensional vector space. (c) Training curve: rapid convergence of the quantum classifier with genetic algorithm.
    }
    \label{fig:binary-dra}
\end{figure}

\textbf{Implementation Details}. For this problem, we will choose $L=7$ and $U_l$ as $R_y(\bar{\theta}_l \cdot \bar{\mathbf{x}})$ for even layers and $R_z(\bar{\theta}_l \cdot \bar{\mathbf{x}})$ for odd layers, where $\bar{\mathbf{x}}$ is the original input data concatenated with a $1$ at the end for bias, and $\cdot$ is the vector dot product. The input data, classically extracted features, and the training curves are summarized in Figure
\ref{fig:binary-dra}

\subsection{GA-based adversarial attacks}
Since the end-to-end classification pipeline is non-differentiable, conventional gradient-based adversarial attacks, such as the Fast Gradient Sign Method (FGSM), are infeasible for our quantum classifier. Even when an attacker has access to the model’s architecture and trained parameters, crafting adversarial samples remains highly challenging without gradient information. Consequently, alternative attack strategies must be considered. Viable methods include: 1) Evolutionary algorithms (EA), such as the Genetic Algorithm (GA)~\cite{holland1992genetic}; 2) Decision-based attacks, which rely on model decisions to iteratively refine adversarial samples; 3) Score-based attacks, leveraging confidence scores to guide adversarial perturbations; and 4) Query-based attacks, which generate adversarial examples by querying the model with various inputs.

In this paper, we employ a GA-based adversarial image generator to attack the classifier. GA, a class of evolutionary algorithms inspired by natural selection~\cite{holland1992genetic}, is particularly suited for complex optimization problems that are intractable using traditional methods. In the case of our proposed quantum classifier, where the optimization objective is non-trivial, GA increases the likelihood of finding a global optimum without relying on gradient information. Despite its effectiveness, GA-based optimization has known limitations that may impact both the efficacy and efficiency of the adversarial attack, such as:

\begin{itemize}
    \item Premature convergence: although GA is designed to reach global optima through mutation, there is still a risk of early stagnation in local optima. This occurs when the fitness scores of certain individuals closely approximate the true global optimum, leading to reduced genetic diversity and ineffective exploration in subsequent iterations.
    \item Non-deterministic behavior: GA relies on stochastic processes, including \textit{initialization, mutation, and crossover}, making its optimization outcomes inherently unpredictable. This variability results in inconsistencies in attack performance, as different runs may yield significantly different adversarial samples.
    \item Computationally expensive: GA-based optimization is resource-intensive, as each iteration involves initializing a large population of candidate solutions and evaluating their fitness. Additionally, repeated evaluations over multiple generations amplify the computational burden, creating a bottleneck that slows down optimization, particularly for high-dimensional problems.
\end{itemize}

To formulate the problem, we aim to generate an adversarial sample $\bar{x}_{adv}$ from an input image $\bar{x}$ with ground-truth label $y$, such that $\bar{x}_{adv}$ is misclassified as a target class $y_{adv} \neq y$. The GA-based optimizer begins by initializing a population $P$ of $N$ randomly generated images, where each pixel value $x_{ij}$  is uniformly sampled $ U(0,255)$. The algorithm then selects the top $k$ candidates with the highest fitness scores for mutation. During mutation, all samples except the best-performing individual undergo alterations with probability $p$. After mutation, the algorithm performs crossover by randomly pairing mutated parent samples and generating offspring, ensuring the final population remains of size $N$. To prevent indefinite execution due to non-convergence, the evolution process terminates after a fixed number of iterations or when the top 10 candidates remain unchanged for consecutive iterations—whichever occurs first. The adversarial sample is selected as the individual with the highest fitness score in the final population. The fitness function, defined in Equation~\ref{eq:fitness}, balances two objectives: maximizing the probability $p_{adv}$ of misclassification into $y_{adv}$, while minimizing the root mean squared error (RMSE) between $\bar{x}$ and $\bar{x}_{adv}$. The weights $w_0$ and $w_1$ control this trade-off—higher $w_0$ prioritizes fooling the classifier, while higher $w_1$ emphasizes imperceptible perturbations, making $\bar{x}_{adv}$ visually similar to $\bar{x}$. The full GA-based attack procedure is outlined in Algorithms~\ref{alg:ga_attack} and \ref{alg:helpers}.

\begin{equation}\label{eq:fitness}
    F = w_0 \cdot p_{adv} - w_1 \cdot \text{RMSE}(\bar{x}, \bar{x}_{adv})
\end{equation}

\begin{algorithm}
\KwData{classifier $\mathcal{M}$, input image $\bar{x}$, mutation rate $p$, population size $N$}
\SetKwFunction{Uniform}{\textsc{Uniform}}
\SetKwFunction{Fitness}{\textsc{Fitness}}
\SetKwFunction{Mutate}{\textsc{Mutate}}
\SetKwFunction{Crossover}{\textsc{Crossover}}
\SetKwFunction{Normal}{\textsc{Normal}}
\SetKwFunction{TopCandidates}{\textsc{TopCandidates}}

\SetKwProg{Fn}{Function}{:}{}
\SetKw{And}{and} \SetKw{Not}{not} \SetKw{In}{in} \SetKw{Or}{or}
\For{$i \leftarrow 1,\cdots, N$}{
    \For{$j \leftarrow 1, \cdots, numPixels$}{
        $P[i][j] \leftarrow \Uniform(0, 255)$\;
    }
}
 \While{$nIters < maxIters$}{
    $P, S \leftarrow \TopCandidates(P, k)$\;
    $P^* \leftarrow \Mutate(\Crossover(P^*, N), p_m)$\;
    $S^* \leftarrow \Fitness(P^*)$\;
    $\bar{x}_{adv}, bestScore \leftarrow \TopCandidates(P^*, 1)$\;
    $C_{prev}, S_{prev} \leftarrow \TopCandidates(P, 10)$\;
    $C, S \leftarrow \TopCandidates(P^*, 10)$\;
    \If{$C_{prev} = C \text{ for } I \text{ consecutive iterations}$}{
        break\;
    }
 }
 \SetKwInOut{Output}{Output}
 \Output{$\bar{x}_{adv}$}
  \caption{GA-based adversarial attack}
 \label{alg:ga_attack}
\end{algorithm}

\begin{algorithm}
\SetKwFunction{Uniform}{\textsc{Uniform}}
\SetKwFunction{Fitness}{\textsc{Fitness}}
\SetKwFunction{Mutate}{\textsc{Mutate}}
\SetKwFunction{Crossover}{\textsc{Crossover}}
\SetKwFunction{Normal}{\textsc{Normal}}
\SetKwFunction{TopCandidates}{\textsc{TopCandidates}}
\SetKwProg{Fn}{Function}{:}{}
\SetKw{And}{and} \SetKw{Not}{not} \SetKw{In}{in} \SetKw{Or}{or}

\Fn{\Crossover{$population, N$}}{
    $nChildren \leftarrow N-\text{size}(population)$\;
    $children \leftarrow \{\}$\;
    \While{$\text{size}(children) < nChildren$}{
        sample parent pair $\{pr_0, pr_1\}$ from $population$\;
        $child_0 \leftarrow pr_0, child_1 \leftarrow pr_1$\;
        \For{$h \in imageHeight$}{
            \For{$w \in imageWidth$}{
                $m \sim \text{Bern}(0.5)$\;
                \If{$m = 1$}{
                    exchange $child_0^{(h,m)}$ with $child_1^{(h,m)}$\;
                }
            }
        }
        $children.\textsc{Append}(child_0)$\;
        $children.\textsc{Append}(child_1)$\;
    }
    $population.\textsc{Extend}(children)$\;
    \Return $population$\;
 }
\Fn{\Mutate{$population, p_m$}}{
$topC, topS \leftarrow \TopCandidates(population, 1)$\;
\For{$P_i \in population$}{
    \If{$p_i \neq topC$}{
        $randNum1 \leftarrow \Uniform(0,1)$\;
        \If{$randNum1 < p_m$}{
            $pixels \leftarrow \text{ randomly sample 10\% of pixels from image }P_i$\;
            \For{$pix \in pixels$}{
                $pix \leftarrow pix + \mathcal{N}(0,1)$\;
            }
        }
    }
    \Return $population$\;
}
 }
 \Fn{\TopCandidates{$population, k$}}{
 $scores = \Fitness(population)$\;
 $C, S \leftarrow \text{select top $k$ candidates $C$ with corresponding scores $S$}$\;
 \Return $C, S$\;
 }
 \caption{Helper functions for GA-based attacks}
  \label{alg:helpers}
 \end{algorithm}

 \textbf{Implementation Details.} To attack the quantum classifier, we assume a complete black-box setting where no internal details of the model—such as the data processing pipeline, dimensionality reduction, or classifier parameters—are accessible or modifiable after training. The GA-based adversarial image generator takes a benign image as input and optimizes it using only the output probabilities from the classifier. The objective is to generate an adversarial sample that maximizes the classifier’s confidence in the target adversarial class $y_{adv}$ while deviating from its original ground-truth label $y$. This objective corresponds to the first term of the fitness function $F$ in Equation~\ref{eq:fitness}. For our experiments, we use a population size of 200 and a maximum of 500 iterations. The weight $w_0$ is fixed at 1, while $w_1$ is varied across 0.9, 1.2, and 1.5 to evaluate the trade-off between attack effectiveness and imperceptibility. Due to the computational cost of generating adversarial images, each trial is limited to producing 100 samples.

\textbf{Attack Outcomes.} Fig. \ref{fig:resilience-vs-depolar}b illustrates an example of benign and adversarial images for digits 0 and 1. It is observable that the adversarial image retains the original image structure, with some ``grains'' in the body of the digits as well as in the surrounding pixels.

% \begin{figure}
% \centering     %%% not \center
% \subfigure[Benign]{
% \includegraphics[width=0.15\textwidth]{figures/digit_benign.png}}
% \subfigure[Adversarial]{
% \label{fig:b}\includegraphics[width=0.15\textwidth]{figures/digit_adv.png}}
% \caption{A benign and adversarial versions of digit 1}
% \label{fig:gen_image}
% \end{figure}

\subsection{Defense with convolutional quantum classifier}
\label{sec:CQC-arch}

% \subsection{Adversarial attacks against quantum classifier}

% TODO: fill in architecture description.

% With the adversarial attack, we are able to maximally perturb the classifier by injecting minimum amount of noise. The accuracy on the physical Ion-trap based classifier dropped from 98\% to 45\%.

% \begin{figure}[ht]
%     \centering
%     \subfigure[\hspace{0mm} ]{
%         \includegraphics[width=0.45\textwidth]{figures/mnist_01_attacked.png}
%         \label{fig:fig2a}
%         }
%     \vspace{0 mm}
%     \subfigure[\hspace{0mm} ]{
%         \includegraphics[width=0.45\textwidth]{figures/mnist_01_attacked_pca.png}
%         \label{fig:fig2b}
%         }
%     \caption{(a) Perturbed data: illustration of attack using minimal perturbation. (b) PCA projection: projection of (a) down to a 2 dimensional vector space.
%     }
%     \label{fig:fig2}
% \end{figure}

% \subsection{Adversarial defense against attacks}

As we can see from Fig. \ref{fig:binary-dra-b}, the main point of attack lies in the bottleneck layer from PCA projection. Rather than using a PCA as a feature exactor for the dataset, we use a family of end-to-end architectures, analogous to classical convolutional neural networks, that automatically learns the features from raw images.

\textbf{Problem definition.} Similar to classical image recognition, we define our input data to be a 2-dimensional tensor $X_{ij} \in [0,1]$ with $1\leq i \leq H$ and $1\leq j \leq W$ where $H$ and $W$ are the height and width of the input image, and the tensor represents the pixel values of an input image. The task is to classify a given input image into $N$ distinct classes.

\textbf{Architecture.} The architecture family we used in our experiment consists of the following. The model alternates between tensors $T_l,1\leq l\leq L+1$ and grids of qubits $|\phi_{l,h,w}\rangle,1\leq l\leq L$, with $T_0$ and $T_{L+1}$ being the input and output respectively. At each layer $1\leq l\leq L$ of the classifier, a grid of $h_l\times w_l$ qubits is initialized to ground state. For each qubit $|\phi_{l,h,w}\rangle$, we \textit{upload} a patch, $T_{l,p\pm dp,q\pm dq}$, to the qubit of interest. $T_{l,p,q}$ is the center of $|\phi_{l,h,w}\rangle$'s receptive field and $dp,dq$ are the size of the receptive field \cite{lecun2015deep}. Lastly, we project the qubit to a pre-defined state, $\langle \psi_{l,h,w}|$, for further computation. Putting the above together, the upload and projection formula for $|\phi_{l,h,w}\rangle$ can be written as:

\begin{align}
    \langle\psi_{l,h,w}|\phi_{l,h,w}\rangle = \langle\psi_{l,h,w}| \prod_{k=1}^K U_k(\sum_{-dp\leq i\leq dp,-dq\leq j\leq dq} \theta_{l,k,i,j} T_{l,p+i,q+j}) |0\rangle \
    \label{equation:cqc-single-qubit}
\end{align}

\begin{figure}[ht]
    \centering
    \includegraphics[width=0.95\textwidth]{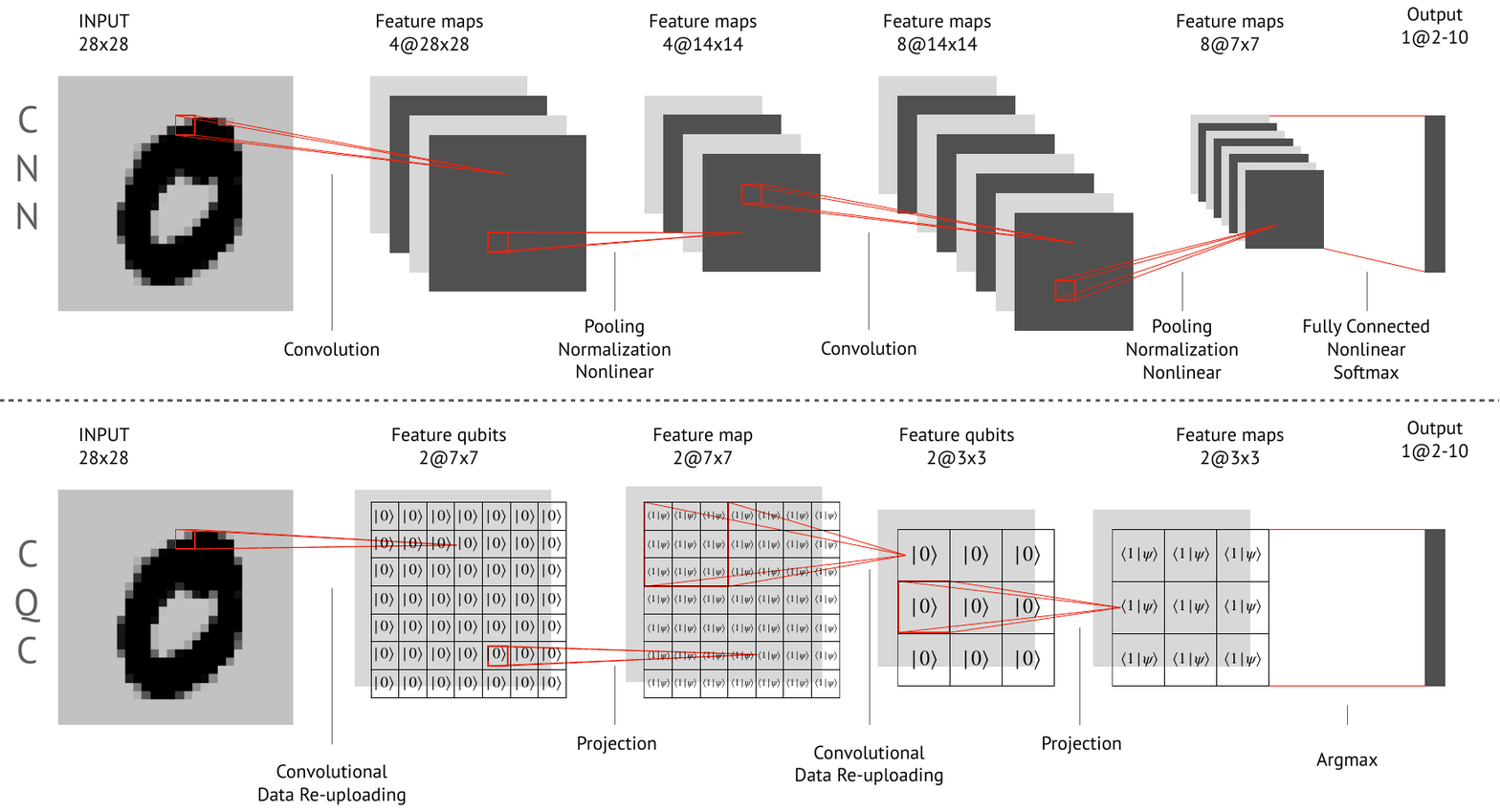}
    \caption{Top: simple classical convolutional neural network (CNN) consisting of Conv2D, Pooling, Normalization, Nonlinear, Fully Connected, and Softmax layers. The output vector can be of length 2 to 10 depending on the number of unique labels we are trying to classify. Bottom: example architecture of a convolutional quantum classifier (CQC) via data re-uploading. From left to right: 1) take a sample image, 2) initialize 2 grids of 7x7 qubits; for each small patch of the image (e.g. 3x3 patch), apply the data re-uploading algorithm to a qubit in ground state $|0\rangle$ using some (e.g. Lx3x3) trainable parameters, 3) project the grid of qubits into a fixed state (e.g. $|1\rangle$), 4) repeat until we have a small enough subspace; perform qubit readout and apply softmax (optionally with a fully connected layer in front).
    }
    \label{fig:dra-cqc}
\end{figure}

\textit{Example} Fig. \ref{fig:dra-cqc} illustrates a possible DRA-CQC architecture with $L=2$. $T_1$ is 28x28, $T_2$ is 7x7, and $T_3$ is 3x3. $|\phi_1\rangle$ is 7x7 and $|\phi_2\rangle$ is 3x3. $dp,dq$ for each layer is fixed as 1 (thus 3x3 receptive field). $\langle\psi_{l,h,w}|$ is chosen to be $\langle 1|$ for every qubit. And since this is a binary classification problem, only the first 2 qubits are considered for classification while the remaining 7 are treated as Ancilla qubits.

\textit{Trainable Parameters.} In the above architecture, the only trainable parameters (weights) of our classifier are the $\theta_{l,k,i,j}$, and it is shared amongst the qubits of the same layer. This is an important choice as it allows the trained classifier to pick out common 2-dimensional sub-features of different image patches, analogous to the filter weights of convolutional neural networks \cite{lecun2015deep}.

\textit{Hyper-parameters.} We have a few hyper-parameters in the above architecture: the number of layers, $L$, the depth of each layer $K$, the width and height of each layer's receptive field $dp,dq$, and the pre-defined projection state at each layer $\langle\psi_{l,h,w}|$.

\subsection{Experimental verification with ion-trap based NISQ device}
\label{sec:exp-nisq}

The experimental setup of the quantum-classical hybrid classifier is structured into three primary functional layers:  the Quantum Processing Unit (QPU), the middleware, and the Classical Processing Unit (CPU). Notably, the QPU and middleware largely build upon previous work \cite{dutta2024trainability}, though the classical processing layer has been enhanced to specifically cater to the new architectural requirements.

At the core of the QPU is an ion-trap architecture based on $^{138}$Ba$^+$ ions, which serves as the platform for qubit initialization, manipulation, and measurement. Ions are confined in a linear Paul trap, where axial and radial confinement frequencies of approximately \( 2\pi \times 0.5 \, \text{MHz} \) and \( 2\pi \times 1.5 \, \text{MHz} \), respectively, ensure the stability of the ions' motion. The system utilizes a magnetic field of 0.72 mT generated by low-temperature coefficient Sm$_2$Co$_{17}$ permanent magnets, to define the quantization axis along the desired direction. The quantization axis is oriented at an angle of approximately 45° from the trap axis, consistent with the system architecture described in prior work by Dutta et al. \cite{dutta2022single, dutta2024trainability},where the qubit states of interest correspond to the ground and metastable excited states, specifically the \( {\rm S}_{\frac{1}{2},-\frac{1}{2}} \) and \( {\rm D}_{\frac{5}{2},-\frac{1}{2}} \) states. Notably, the transition frequency between these states is first-order insensitive to magnetic field fluctuations, thereby significantly mitigating decoherence effects and enhancing both the coherence time and operational fidelity of quantum gates.

A crucial element of the QPU is the narrow-linewidth \( 1762 \, \text{nm} \) laser, which is stabilized to an ultra-stable cavity, ensuring a linewidth of approximately 100 Hz~\cite{yum2017optical, dutta2020single}. This level of spectral precision is critical for implementing high-fidelity single-qubit gates with minimal dephasing. However, magnetic field noise and residual timing jitters in gate pulses remain non-negligible sources of dephasing. To control the phase, frequency, and amplitude of the laser pulses during gate implementation, the electro-optic (EO) and acousto-optic (AO) layers are employed. These layers, governed by a combination of stable radio-frequency generators and amplifiers, provide the necessary hardware control to perform quantum operations with high fidelity. The radio-frequency generators utilize direct digital synthesizers (DDS), such as the AD9958 chip, which offer precise control over frequency, phase, and amplitude within the range of \( 20-250 \, \text{MHz} \), with resolutions of 32 bits, 16 bits, and 10 bits, respectively.

The middleware is pivotal in coordinating the sequence of quantum operations and ensuring the accuracy of the quantum state measurements. Field programmable gate arrays (FPGAs), based on the Altera Cyclone V chip, are responsible for managing the algorithmic time sequence of the quantum operations, as well as collecting the final state measurements of the qubit. This interaction between the quantum processor and classical control layers enables the QPU to perform the necessary quantum gate operations, with a focus on minimizing errors and optimizing fidelity.

Prior to each execution cycle, the quantum processor undergoes an initialization and cooling sequence to prepare the qubit for gate operations. Doppler cooling, achieved via a fast dipole transition at \( 493 \, \text{nm} \)combined with a repump laser at \( 650 \, \text{nm} \), brings the ion to the Lamb-Dicke regime. The cooling beam is meticulously aligned along the trap axis to efficiently address all motional modes of the trapped ions, ensuring that coherence is maintained throughout the computational sequence. This precise overlap of the cooling beam with motional modes is critical for stabilizing the qubit state and maintaining high-fidelity gate operations.

\backmatter

% \bmhead{Supplementary information}

% If your article has accompanying supplementary file/s please state so here. 

% Authors reporting data from electrophoretic gels and blots should supply the full unprocessed scans for key as part of their Supplementary information. This may be requested by the editorial team/s if it is missing.

% Please refer to Journal-level guidance for any specific requirements.

% \bmhead{Acknowledgements}

% Acknowledgements are not compulsory. Where included they should be brief. Grant or contribution numbers may be acknowledged.

% Please refer to Journal-level guidance for any specific requirements.

\section*{Declarations}

\begin{itemize}
\item Funding\\
This research is supported by the National Research Foundation, Singapore, and A*STAR under its Quantum Engineering Programme (NRF2021-QEP2-02-P10 and NRF2021-QEP2-01-P01).
\item Conflict of interest/Competing interests (check journal-specific guidelines for which heading to use)\\
We declare that there is no conflict of interest for this work.
\item Data, and code availability\\
The data and code supporting the findings of this study are available from the corresponding author upon reasonable request.
\end{itemize}

% \begin{appendices}

% \section{Section title of first appendix}\label{secA1}

% An appendix contains supplementary information that is not an essential part of the text itself but which may be helpful in providing a more comprehensive understanding of the research problem or it is information that is too cumbersome to be included in the body of the paper.




% \end{appendices}


\bibliography{sn-article}% common bib file

\end{document}